\documentclass{svproc}
\usepackage{url}

\usepackage{graphicx}
\usepackage{epsfig}
\usepackage{float}
\usepackage{amsmath}
\usepackage{amssymb}

\begin{document}
\mainmatter
\title{Poincar\'e return maps in neural dynamics: three examples}
\titlerunning{Poincar\'e return maps in neural dynamics}

\author{Marina L. Kolomiets \inst{1} \and Andrey L. Shilnikov \inst{2}}
\authorrunning{Marina Kolomietz \and Andrey Shilnikov}
\tocauthor{Marina Kolomietz and Andrey Shilnikov}

\institute{Department of Mathematics, Academy of Agricultural Sciences,\\ 
Nizhniy Novgorod 603107, Russia
\and
Neuroscience Institute, Department of Mathematics and Statistics,\\ Georgia State University,\\ Atlanta, Georgia 30303, USA,\\
\email{ashilnikov@gsu.edu},\\ WWW home page:
\texttt{https://labs.ni.gsu.edu/ashilnikov/}
}

%

\maketitle
\begin{abstract}
Understanding of the onset and generic mechanisms of transitions between distinct patterns of activity in realistic models of individual neurons and neural networks presents a fundamental challenge for the theory of applied dynamical systems. We use three examples of slow-fast neural systems to demonstrate a  suite of new computational tools to study diverse neuronal systems. 
\end{abstract}

\section{Introduction}

Most neurons demonstrate oscillations of the membrane potential either 
endogenously or due to external perturbations. Deterministic description of primary
oscillatory activities, such as tonic spiking and bursting, of neuronal dynamics is 
based on models following the Hodgkin-Huxley formalism \cite{Hodgkin1952}.
Mathematically, such conductance based models belong to a special class of dynamical
systems with at least two distinct time scales, the so-called slow -- fast
systems \cite{Arnold1994,Bertram1995,Izhikevich2007,Jones1994,Rinzel1985,Rinzel1998,Rinzel1995}. 
Bursting is a manifestation of slow--fast dynamics  possessing subcomponents operating at distinct time scales. Neural bursting is a modular activity composed of various limiting branches, corresponding to oscillatory and equilibrium regimes of the fast subsystem, and connected by transients between them. 
Using the common mathematical we can 
better understand the basic onset of bursting oscillations in models of individual and coupled neurons.
The study of mechanisms of bursting and its transformations requires  nonlocal
bifurcation analysis, which is based on the derivation and further examination of
Poincar\'e return maps.

\section{Hodgkin-Huxley type model of a leech heart interneuron}
 
 Our first example is  the ``reduced''  model of
 heart interneuron model \cite{Shilnikov2005a,Shilnikov2005b,Channell2007a,Channell2007} derived through the Hodgkin-Huxley gated variables formalism \cite{Hodgkin1952} that not every mathematician may be familiar with. Its equations do look too detailed and overwhelming: 
\begin{eqnarray}
&& \displaystyle C\,\frac{dV}{dt} =  -I_{\mathrm{Na}}-I_{\mathrm{K2}}+I_{\mathrm{L}}-I_{\mathrm{app}}-I_{\rm syn} ,\\
&& I_{L}=\bar g_{\mathrm{L}}\,(V-E_{\mathrm{L}}),\qquad   I_{\mathrm{K2}}={\bar g}_{\mathrm{K2}}\,m_{\mathrm{K2}}^2(V-E_{\mathrm{K}}),  \nonumber   \\
&& I_{\mathrm{Na}}={\bar g}_{\mathrm{Na}}\,m^3_{\mathrm{Na}}\,h_{\mathrm{Na}}\,(V-E_{\mathrm{Na}}),\quad m_{\mathrm{Na}}=m^\infty_{\mathrm{Na}}(V),  \nonumber  \\
&& \tau_{\mathrm{Na}}\, \displaystyle   \frac{dh_{\mathrm{Na}}}{dt} = h^\infty_{\mathrm{Na}}(V)-h,~~~
   \tau_{\mathrm{K2}} \displaystyle   \frac{dm_{\mathrm{K2}}}{dt} = m^\infty_{\mathrm{K2}}(V)-m_{\mathrm{K2}}   \nonumber ,
\label{biot}
\end{eqnarray}
where $C=0.5$~nF is the membrane capacitance; $V$ is the membrane potential; $I_{\mathrm Na}$ is the fast voltage gated sodium current with slow inactivation $h_{\mathrm{Na}}$ and fast activation $m_{\mathrm{Na}}$; $I_{\mathrm{K2}}$ is the persistent potassium current with activation $m_{\mathrm{K2}}$; $I_\mathrm{L}$ is leak current and $I_{\mathrm{app}}$ is a constant polarization or external applied current. The maximal conductances are ${\mathrm{\bar g}}_{\rm K2}=30$nS, ${\bar g}_{\rm Na}=200$nS and  $\mathrm{g}_{\rm L}=8$nS, and the reversal potentials are $\mathrm{E}_{\rm Na}=0.045$~V, $\mathrm{E}_{\rm K}= -0.070$V and $E_{\mathrm{L}}=-0.046$V.
The time constants of gating variables are $\tau_{\rm K2}=0.25$~sec and $\tau_{\rm Na}=0.0405~$sec.  The steady state values of gating variables, $h^\infty_{\mathrm{Na}}(V)$, $m^\infty_{\mathrm{Na}}(V)$, $m^\infty_{\mathrm{K2}}(V)$,
are given by the following sigmoidal functions:
 \begin{equation}
\begin{array}{rcl}
 h^\infty_{\mathrm{Na}}(V)&=&[1+\exp(500(0.0333-V))]^{-1}\\
 m^\infty_{\mathrm{Na}}(V)&=&[1+\exp(-150(0.0305-V))]^{-1}\\
 \quad m^\infty_{\mathrm{K2}}(V)&=&[1+\exp{(-83(0.018-V+\mathrm{V^{shift}_{K2}}))}]^{-1}.
\end{array}
\end{equation}\label
The quantity $\mathrm{V^{shift}_{K2}}$ is a genuine bifurcation parameter for this model: it is the deviation from
experimentally averaged voltage value $V_{\rm 1/2}=0.018$V corresponding to semi-activated potassium channel, i.e. $m^{\infty}_{\mathrm{K2}}(0.018)=1/2$.  Variations of $\mathrm{V^{shift}_{K2}}$  move the slow nullcline $\frac{dm_{\mathrm{K2}}}{dt}=0$ in the $V$-direction in the 3D phase, see Fig.~1. Due to the disparity of the time constants of the phase variables, the fast-slow system paradigm is applicable to system~(1): its first two differential equations form a fast subsystem, while
the last equation is the slow one. The dynamics of such a system
are known \cite{fenichel-tikhonov} to be determined by, and
centered around, attracting pieces of the slow motion manifolds
that constitute a skeleton of activity patterns. These manifolds
are formed by the limit sets, such as equilibria and limit cycles,
of the fast subsystem where the slow variable becomes a parameter
in the singular limit.

A typical Hodgkin-Huxley model possesses a
pair of such manifolds \cite{rinzel}: quiescent and tonic spiking,
denoted by $\mathrm{M_{eq}}$ and $\mathrm{M_{lc}}$,
correspondingly. A solution of (\ref{biot}) that repeatedly
switches between the low, hyperpolarized branch of
$\mathrm{M_{eq}}$ and the spiking manifold $\mathrm {M_{lc}}$
represents a busting activity in the model.  
Whenever the spiking manifold $\mathrm {M_{lc}}$ is transient for
the solutions of (1), like those winding around it in
Figs.~\ref{fig2}, the models exhibits regular or chaotic bursting.
Otherwise, the model~(1) has a spiking periodic orbit that has
emerged on $\mathrm{M_{lc}}$ through the saddle-node bifurcation
thereby terminating the bursting activity \cite{sst} or both
regimes may co-exist as in \cite{pre,jcn}. 

\begin{figure}[ht!]
\centering
\includegraphics[width=0.75 \columnwidth]{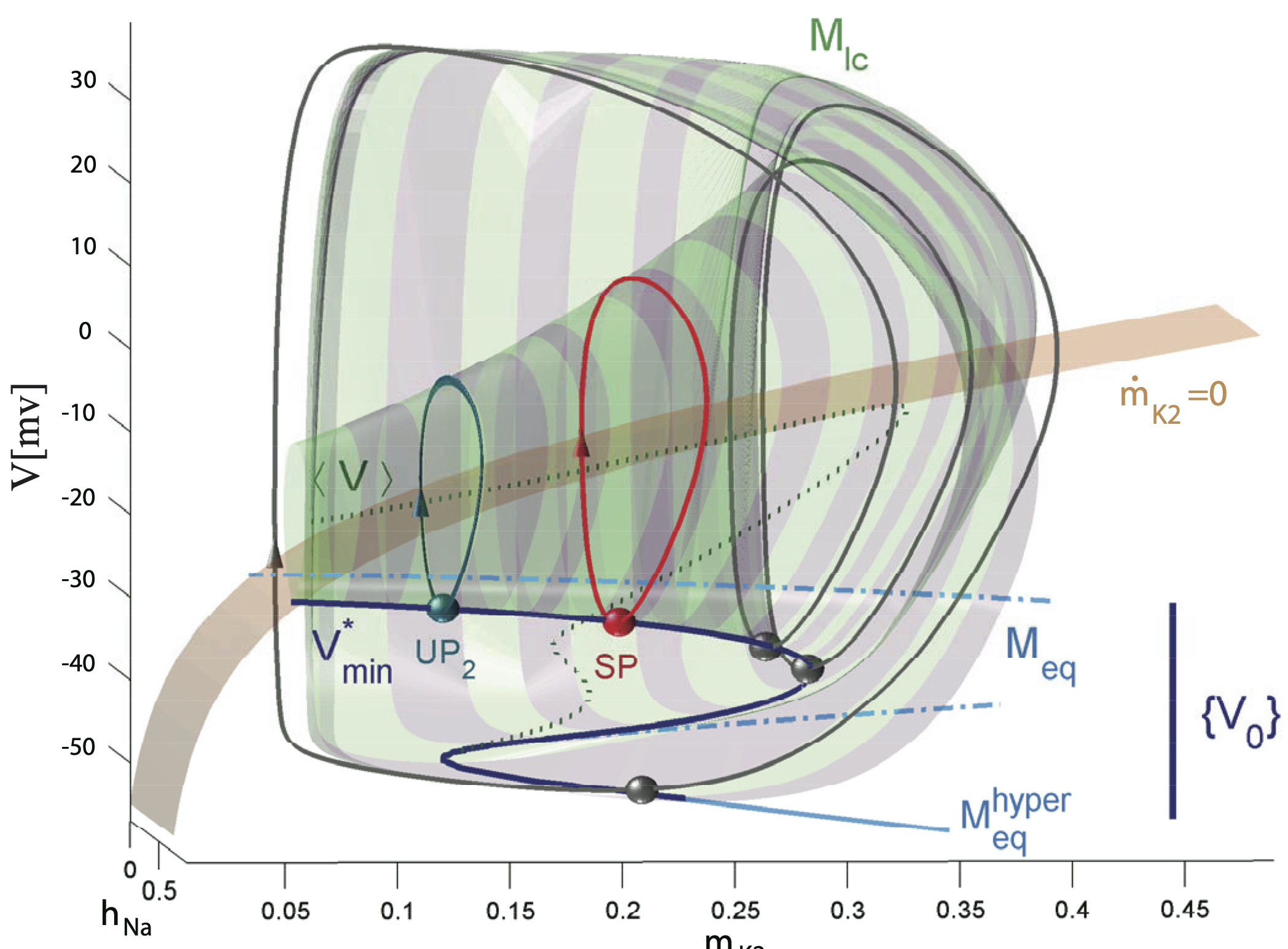}
  \caption{Slow motion manifolds and nullclines of the model (1): the
2D spiking manifold $\mathrm{M_{lc}}$ is foliated by the periodic orbits continued, from the left to the right,  as the parameter $\mathrm{V_{K2}^{shift}}$ is increased from $-0.026$ through $0.0018$.  The space curves $\mathrm{V_{min}}$ and $\langle \rm V \rangle$ are made of minimal and average coordinates of the periodic orbits. $\mathrm{M_{lc}}$ glues to the hyperpolarized
fold of  the quiescent manifold, $\mathrm{M_{eq}}$, comprised of the equilibrium states of (\ref{biot}), where the curve of the averaged
values $\langle \rm V \rangle$ terminates.  An equilibrium state of Eqs.~(1) is the intersection point of $\mathrm{M_{eq}}$ with the slow (yellow) nullcline ${\dot m_{K2}}=0$ for given $\mathrm{V^{shift}_{K2}}$. Also shown (in red) is the curve of the $v$-minimal coordinate values of the periodic orbits making $\mathrm{M_{lc}}$. This curve is used to define the Poincar\'e map taking it onto itself after one revolution around $\mathrm{M_{lc}}$.}\label{fig1}
\end{figure}

To determine what makes the spiking and bursting attractors change their shapes and
stability, we construct numerically a $\mathrm{V_{K2}^{shift}}$-
parameter family of 1D Poincar\'e maps taking an interval of
membrane potentials onto itself. This interval is comprised of the
minimal values, denoted by $(\mathrm{V_0})$, of the membrane
potential on the found periodic orbits foliating densely the
spiking manifold $\mathrm{M_{lc}}$, see Fig.~1. Then, for some
$\mathrm{V^{shift}_{K2}}$-values, we integrate numerically
the outgoing solutions of (\ref{biot}) starting from the initial
conditions corresponding to each ($\mathrm{V_0}$) to find the
consecutive minimum $(\mathrm{V_1})$ in the voltage time series.
All found pairs $\mathrm{(V_{0},\,V_{1})}$ constitute the graph of
the Poincar\'e map for given $\mathrm{V^{shift}_{K2}}$. 

\begin{figure}[ht!]
\centering
\includegraphics[width=0.49 \columnwidth]{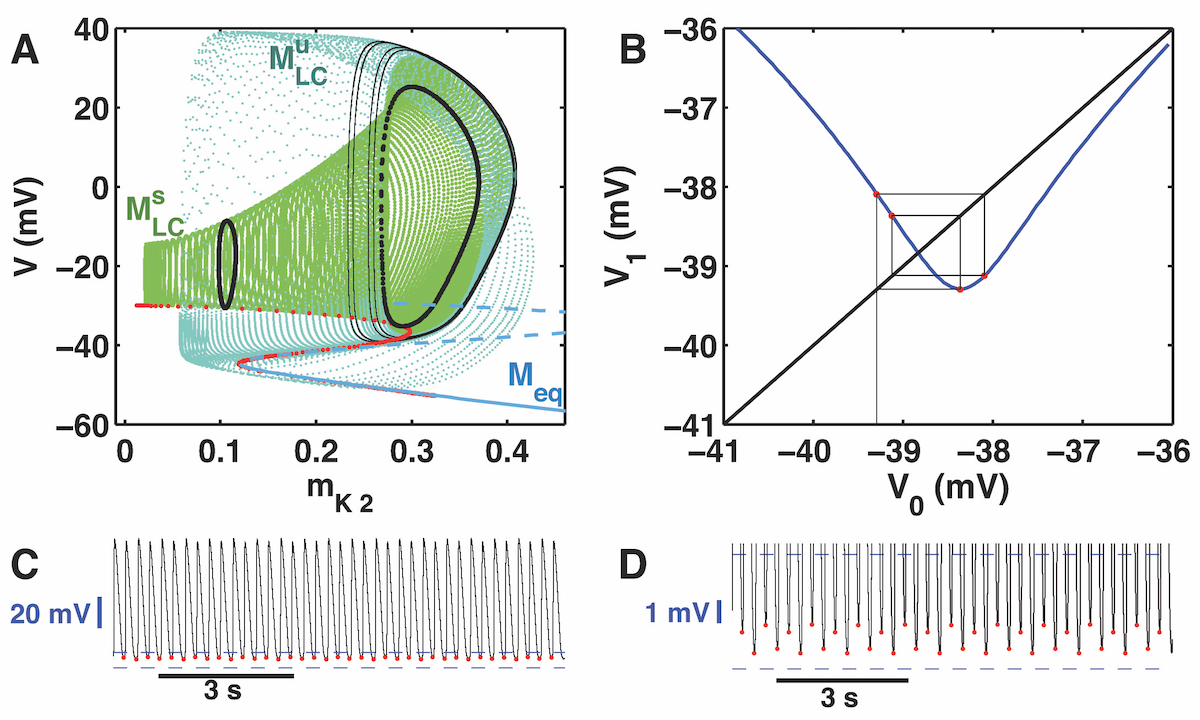}
\includegraphics[width=0.49 \columnwidth]{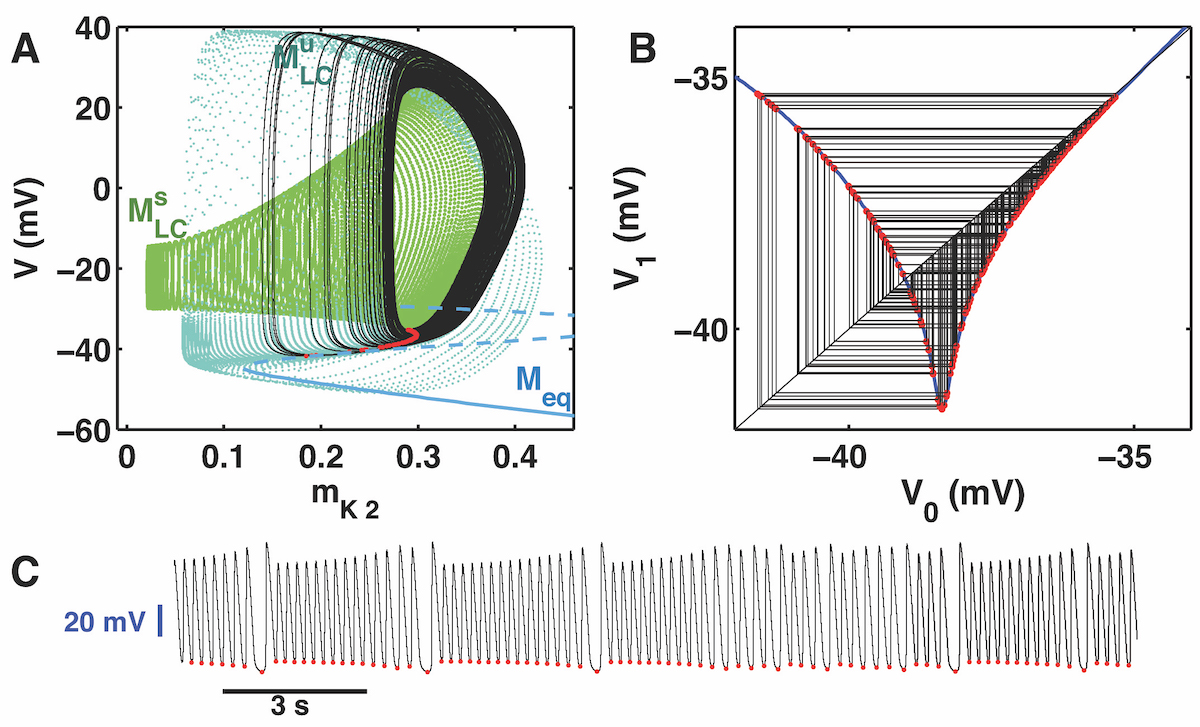}
\includegraphics[width=0.47 \columnwidth]{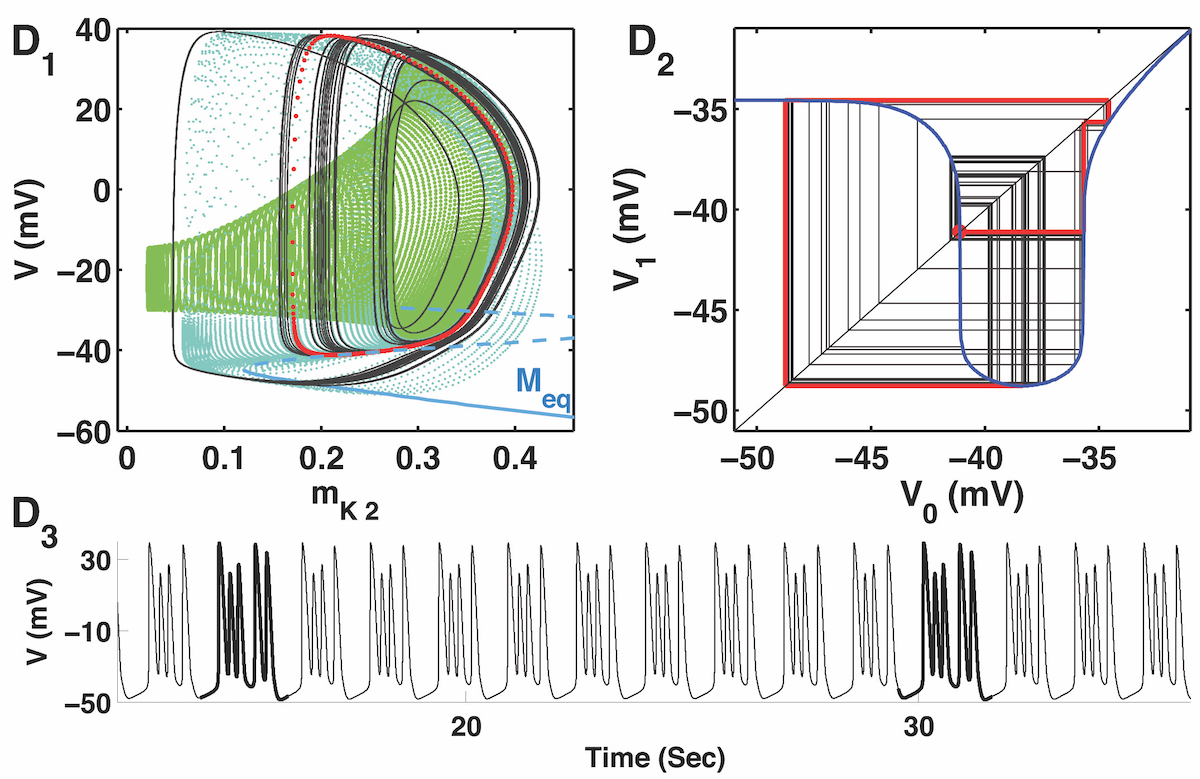}
\includegraphics[width=0.5 \columnwidth]{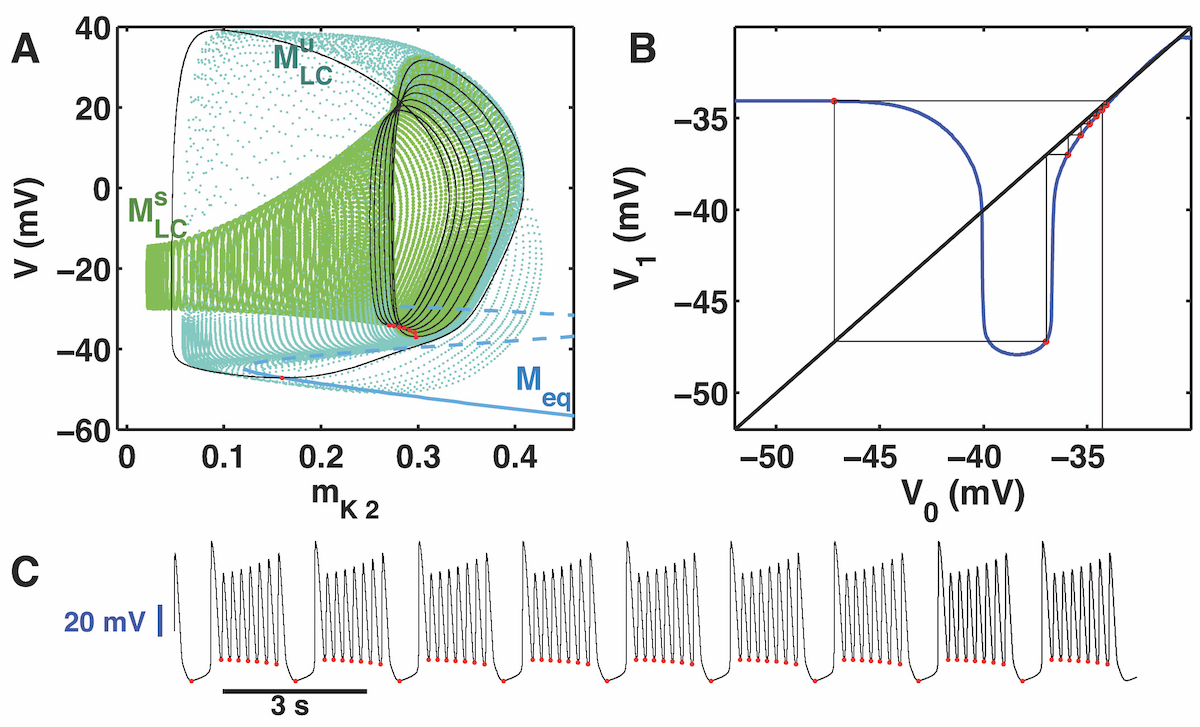}
  \caption{(Top-left) Four $v$-minimums of the stable spiking periodic orbit spiking at $\rm V^{\rm shift}_{\rm K2}=0.0255$ corresponding to the period-4 orbit of the Poincar\'e map. Insets (C) and (D) show the voltage waveforms.(Top-right) Chaotic spiking of the model and in the map at $\rm V^{\rm shift}_{\rm K2}=-0.0254$. (Bottom) Chaotic bursting at the spike adding transition becomes more regularized with a large number of spikes per burst. }\label{fig2}
\end{figure}

Figure~2 is a showcase of such 1D unimodal maps with the distinctive U-shape. 
A fixed point of map would correspond to a single
$\mathrm{V}$-minimum on the periodic orbit on the 2D tonic spiking manifold, while 
period-2 orbit of the map corresponds to the periodic orbit of the model and so forth. 
A bursting orbit with multiple turns around $\mathrm{M_{lc}}$ and switching to and back from  
$\mathrm{M_{lc}}$ is represented by a more complex orbit of a longer period. Moreover, the bursting orbit may become even chaotic at spike adding transition, and as the map reveals that is caused by a homoclinic orbit (red trajectory) of an unstable fixed point corresponding to a saddle periodic orbit of the neural model~(1). The shape of the 1D return map infers that as it becomes steeper with a characteristic cusp shape the model would move into the chaotic regime.

\section{FitzHugh-Nagumo-Rinzel Model}\label{fhr_model}
Our next example is the  FitzHugh-Nagumo-Rinzel (FNR) model which is a mathematical model of an elliptic burster (see Fig.~3(B)); its equations given by \cite{wojcik2011voltage}:
 \begin{equation} \label{fhr}
 \begin{array}{rclcl}
 v^\prime &= & v-v^3/3-w+y+I,\\
 w^\prime &= & \delta(0.7+v-0.8w),\\
 y^\prime &= & \mu(c-y-v).
 \end{array}
\end{equation}
Here,  $\delta=0.08$, $I=0.3125$ is an ``external current'', and we set $\mu=0.002$ determining the pace of the slow variable $y$;
the bifurcation parameter of the model is $c$.
\begin{figure}[ht!]
\centering
\includegraphics[width=0.44 \columnwidth]{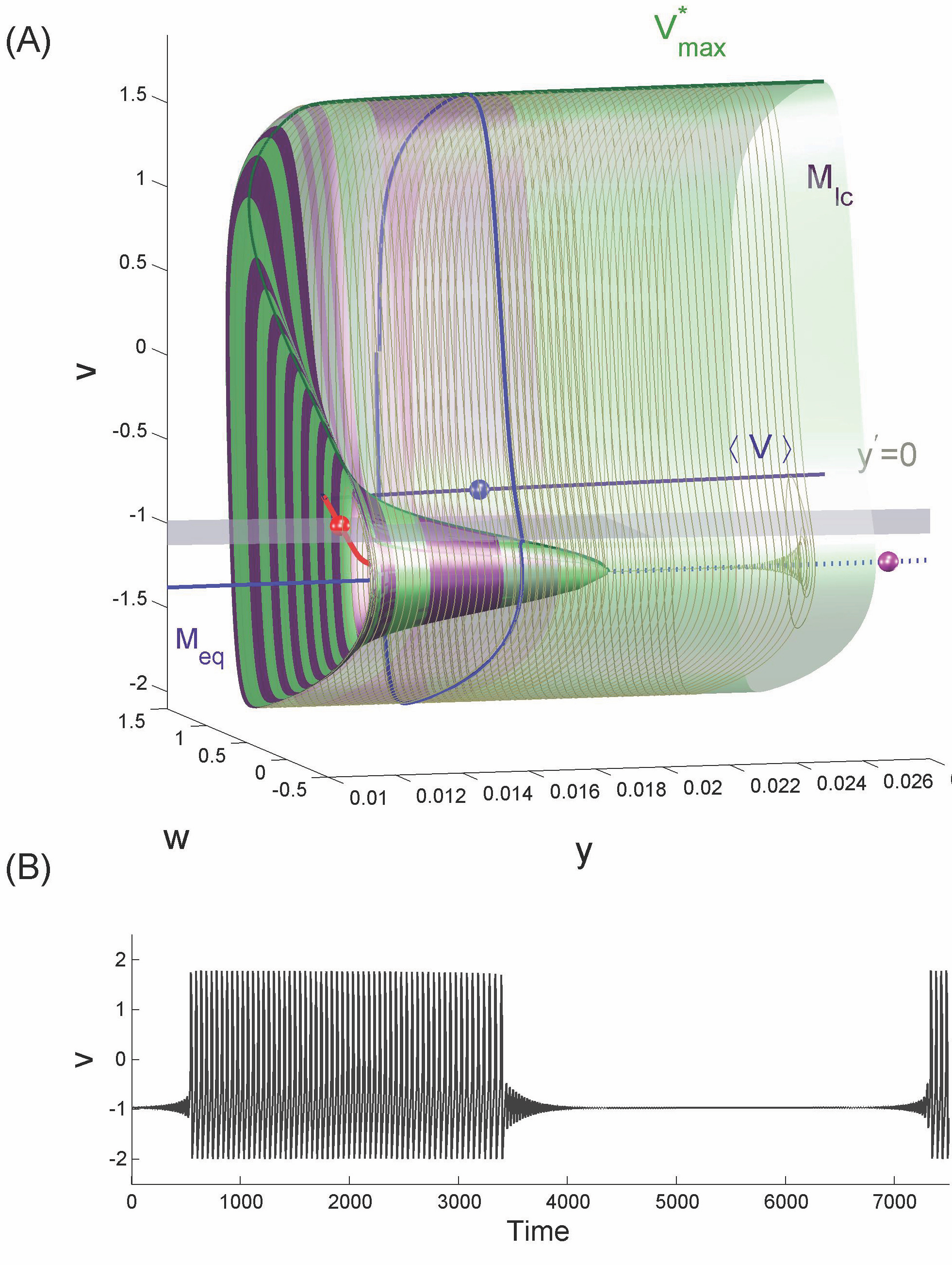}
\includegraphics[width=0.55 \columnwidth]{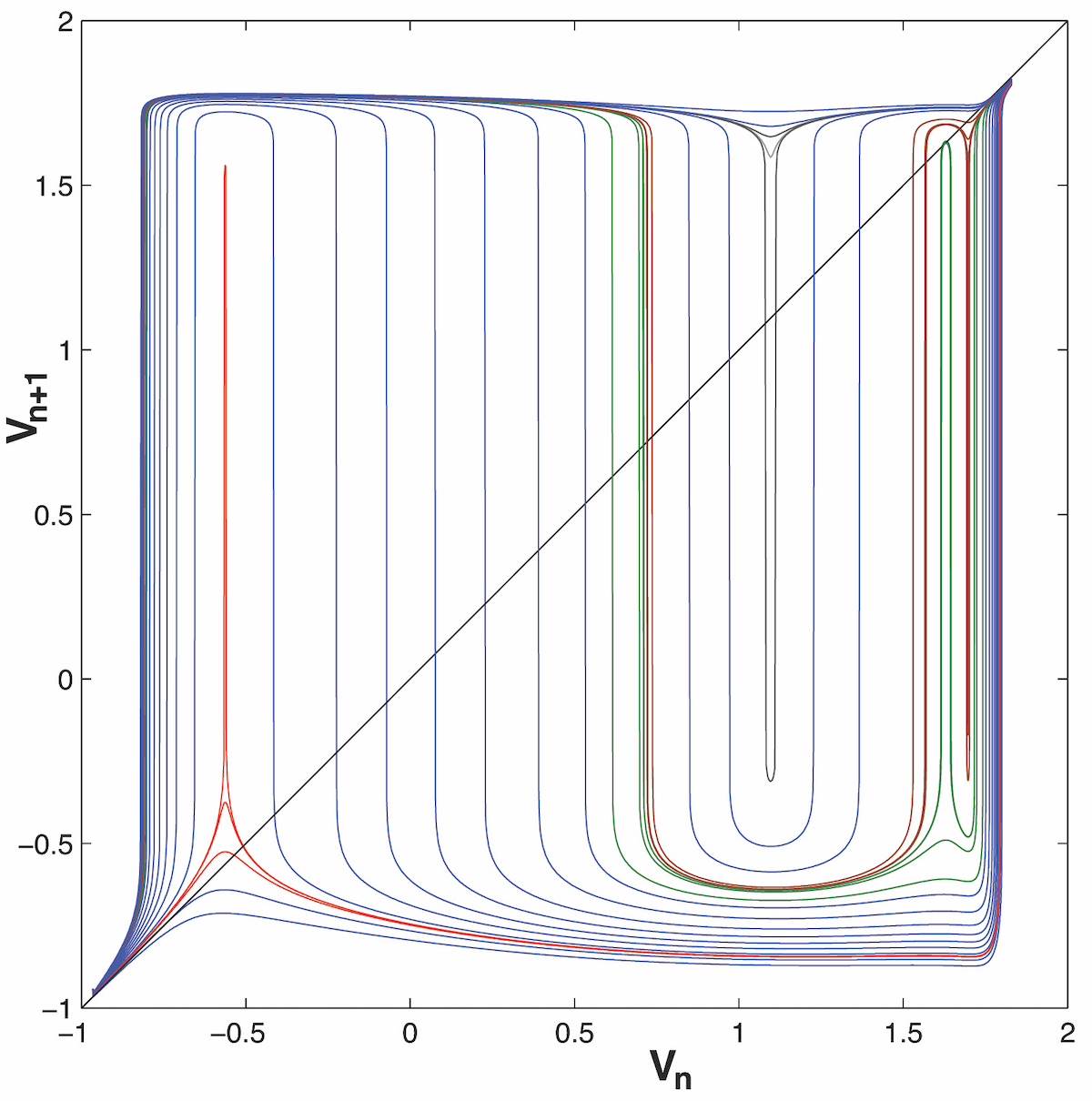}
  \caption{(A) Topology of the tonic spiking, $\mathrm{M_{lc}}$, and quiescent,
$\mathrm{M_{eq}}$, manifolds. The fold on $\mathrm{M_{lc}}$, corresponds to a saddle-node bifurcation where  the stable (outer) and saddle (inner) branches, comprised of periodic orbits, merge. The vertex, where the unstable branch of $\mathrm{M_{lc}}$  collapses at $\mathrm{M_{eq}}$,  corresponds to a subcritical Andronov-Hopf bifurcation. Space curves, labeled by $\mathrm{V^{*}_{max}}$ (in green) and $\langle \mathrm{V^{s,u}} \rangle$ (in blue and red, respectively), correspond to the V-maximal and the averaged, over the period, coordinates of the periodic orbits composing $\mathrm{M_{lc}}$. The plane, $y^\prime=0$, is the slow nullcline, above (below) which the $y$-component
of a solution of the model increases (decreases).  The plane is elevated/lowered 
 as the $c$-parameter is increased/decreased.  (right) The ``continuously'' reshaping family of the 1D Poincar\'e return maps $T: \mathrm{V_n} \to \mathrm{V_{n+1}}$ for the FHN-model at $\mu=0.002$ as $c$ increases from $c=-1$ through $c=-0.55$.  Lower graphs
correspond to quiescence and subthreshold oscillations in the model; upper graphs correspond to tonic spiking dynamics, while the middle graphs describe bifurcations of bursting. An intersection point of a graph with the bisectrix is a fixed point of the map. The stability of the fixed point is determined by the slope of the graph, i.e. it is stable if $|T'|<1$. }\label{fig3}
\end{figure}

The slow variable $y$ becomes frozen when $\mu=0$. The first two fast equations in (\ref{fhr}) compose the FitzHugh-Nagumo fast subsystem model describing a relaxation oscillator, provided $\delta$ is small. This subsystem exhibits either tonic spiking on a stable limit cycle,
or quiescence on a stable equilibrium state for some fixed values of $y$.  Stability loss of the equilibrium state in the fast subsystem gives rise to a stable limit cycle through a sub-critical Andronov-Hopf bifurcation when an unstable limit cycle
collapses into the equilibrium state. The stable and unstable limit cycle emerge in the FNR-model through a saddle-node
bifurcation. Both bifurcations, Andronov-Hopf and saddle-node, are key to the description of an elliptic burster.  Using a traditional slow-fast dissection, one can locate the corresponding branches  of the limit cycle and equilibrium states by varying the frozen $y$-variable  in the extended phase space of the fast subsystem. The topology of  the tonic spiking, $\mathrm{M_{lc}}$, and quiescent, $\mathrm{M_{eq}}$, in the phase space the FNR-model is revealed in Fig.~\ref{fig3}.

\section{1D voltage maps}\label{vmap}

Recall that a feature of a slow-fast system is that its solutions are constrained to stay near the slow-motion manifolds, composed of equilibria and periodic obits of the fast subsystem. If both manifolds are transient for the solutions of the corresponding neuron model, it exhibits a bursting behavior, which is a repetitive alternation of tonic spiking and quiescent periods. Otherwise, the model demonstrates the tonic spiking activity if there is a stable periodic orbit on the tonic spiking manifold, or it shows no oscillations when solutions are attracted to a stable equilibrium state on the quiescent manifold.

The core of the methods is a reduction to, and a derivation of, a low dimensional Poincar\'e return map, with an accompanying analysis of the limit solutions: fixed, periodic and homoclinic orbits, representing various oscillations in the original model. Maps have been actively employed in computational neuroscience, see \cite{Chay1985,Griffiths2006,Shilnikov2003,Shilnikov2004}
and referenced therein. It is customary that such a map is sampled from voltage traces, for example by
singling out successive voltage maxima or minima, or interspike intervals. A drawback of a map generated by time series is a sparseness, as the construction algorithm reveals only a single periodic attractor of a model, unless the latter demonstrates chaotic or mixing dynamics producing a large variety of densely wandering points. 

A new, computer assisted method for constructing  a complete family of Poincar\'e maps
for an interval of membrane potentials for  slow-fast Hodgkin-Huxley models of neurons was proposed in
\cite{Channell2007} following \cite{Shilnikov1993}, see above. Having such maps we are able to elaborate on bifurcations in the question of tonic spiking and bursting, detect bistability, as well examine unstable sets, which are the organizing centers of complex  dynamics in any model. Using this approach we have studied complex bursting
transformations in a leech heart interneuron model and revealed that the cause of complex behaviors at transitions is homoclinic tangles of saddle periodic orbits which can be drastically amplified by small noise \cite{Channell2007a,Channell2009}. Examination of the maps will help us
make qualitative predictions about transitions {\em before} they actually occur in the models.

The construction of the voltage interval maps is a two stage routine. First, we need to accurately single out the slow motion manifold $\mathrm{M_{lc}}$ in the neuronal model using the parameter continuation technique. The manifold is formed by the tonic-spiking periodic orbits as a control parameter in the {\em slow} equation is varied. Recall, that its variations,  raising or lowering the slow nullcline in the phase space of the model, do not alter the fast subsystem and hence do keep the manifold intact. Next a space curve $\mathrm{V^{*}_{max}}$ on $\mathrm{M_{lc}}$ is detected, which corresponds to maximal voltage values of the membrane potentials $V_n$ found on all periodic orbits constituting the tonic spiking manifold, see Fig.~\ref{fig3}.

\begin{figure}[ht!]
\centering
\includegraphics[width=0.8 \columnwidth]{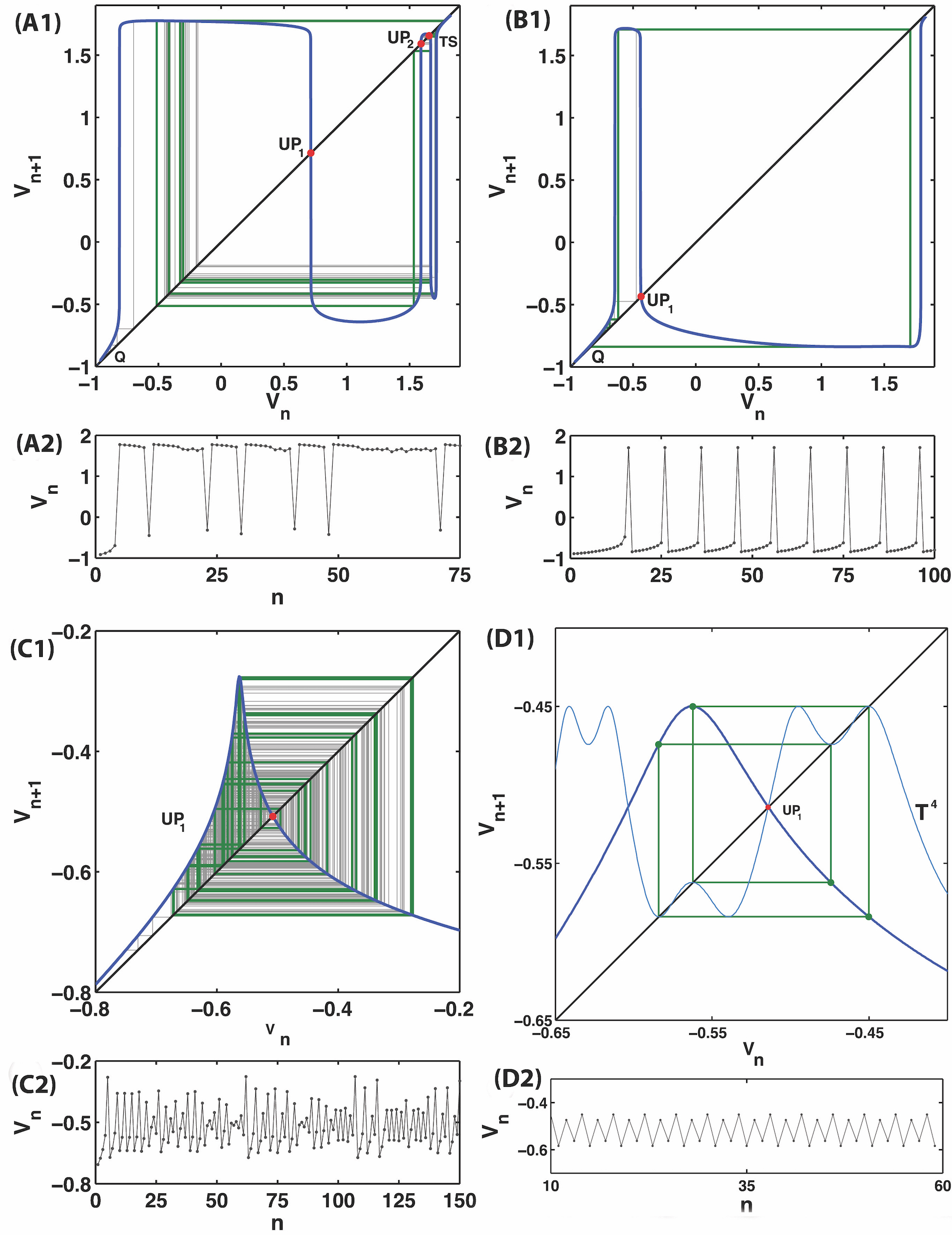}
  \caption{(A1/2) The shape of the 1D Poincar\'e return map reveals the underlying cause of chaotic mixed mode oscillations (MMOs) at the transition from tonic spiking to bursting in the in the FNR-model~(3) that become periodic MMOs with a single burst followed by nine sub-threshold oscillations (B1/2). (C1/2) The unimodal map corresponding to chaotic and period-4 sub-threshold oscillations (D1/2) .}\label{fig4}
\end{figure}

We use this data to further amend the set $\mathrm{\{ V_n \}}$, by integrating the solutions of the model
in the vicinity of each maxima to find the exact locations of the turning points, determined by the condition $\mathrm{V^{\prime}_{max}}=0$. Next, the points defining $\mathrm{\{ V_n \}}$ are employed as the initial conditions to compute outgoing solutions of (\ref{fhr}) that will stay on or close to $\mathrm{M_{lc}}$. The integration is stopped when a successive maximal  value $\mathrm{\{ V_{n+1} \}}$ of the voltage is reached in the voltage trace. Figure~4 demonstrates how the shape of the 1D maps changes in a complex predictable way as the $c$-parameter is varied. One can see from the end points, that the map has initially a stable fixed point at the top-right corner that corresponds to the stable tonic spiking orbit on the outer surface of the 2D manifold $\mathrm{M_{lc}}$ in Fig.~3(left). One can also foresee from the map at the bottom-right corner in Fig.~3(right) the neural model will undergo a cascade of period-doubling bifurcations of sub-threshold oscillations followed by complex mixed-mode oscillations involving sub-threshold ones and bursting. Our predictions are illustrated and confirmed by Fig.~4 that samples 
four characteristic 1D Poincar\'e return maps out of Fig.~3. In it the shape of the 1D Poincar\'e return maps reveals the underlying cause of chaotic mixed mode oscillations (MMOs) at the transition from tonic spiking to bursting in the in the FNR-model~(3) that next become periodic MMOs, and further transition to chaotic and regular sub-threshold oscillations en a route to the quiescent phase in generic elliptic bursters.             

\section{Example 3: 2D recurrent maps in multifunctional 3-cell networks.} 

Many rhythmic motor behaviors such as respiration, chewing, locomotion on land and in water, and heartbeat (in leeches) are produced by networks of cells called central pattern generators (CPGs). A CPG is a neural microcircuit of cells whose synergetic, nonlinear interactions can autonomously generate an array of multicomponent/polyrhythmic bursting patterns of activity that determine motor behaviors in animals, including humans \cite{KCF05,CTMKF07,NSLGK012,CPG,Bal1988,Marder1996,KatzHooper07}. Modeling studies, phenomenologically mathematical and exhaustively computational, have proven useful to gain insights into operational principles of CPGs \cite{Kopell26102004,Ma87,Kopell88,Canavier1994,SKM94,Dror1999,Prinz2003}. Although various models, reduced and feasible, of specific CPGs, have been developed, it remains unclear how the CPGs achieve the level of robustness and stability observed in nature \cite{prl08,Shilnikov2008b,SHWG10,Koch01122011,M12}.
 Understanding the key universal mechanisms of the functional evolution of neural connectivity, bifurcation mechanisms underlying transitions between different neural activities, and accurate modeling of these processes presents opportunity and challenge for applied mathematics in particular and for all computational sciences in general.

Whereas a dedicated CPG generates a single pattern robustly, a multifunctional or polymorphic CPG can flexibly produce distinct rhythms, such as temporally distinct swimming versus crawling locomotions, and alternation of directions of blood circulation in leeches \cite{Calabrese01122011,Kristan2008b,Briggman2008}. Switching between various attractors of a CPG network causes switching between locomotion behaviors. Each attractor is associated with a definite rhythm running on a specific time scale with well-defined and robust phase lags among the constituting neurons. The emergence of synchronous rhythms in neural networks is closely related to temporal characteristics of coupled neurons due to intrinsic properties and types of synaptic coupling, which can be inhibitory, excitatory and electrical, fast and slow \cite{Wojcik2011a,Wojcik2014,pre2012,ftm,Marder1994752}.  

 We developed a computational toolkit for oscillatory networks that reduces the problem of the occurrence of bursting and spiking rhythms generated by a CPG network to the bifurcation analysis of attractors in the corresponding Poincar\'e return maps for the phase lags between oscillatory neurons. The structure of the phase space of the map is an individual signature of the CPG as it discloses all characteristics of the functional space of the network.  Recurrence of rhythms generated by the CPG (represented by a system of coupled Hodgkin-Huxley type neurons \cite{Shilnikov2012}) lets us employ Poincar\'e return maps defined for phase lags between spike/burst initiations in the constituent neurons (Fig.~5) \cite{prl08,Wojcik2011a,Wojcik2014,pre2012,jalil2013}. 
  Forward trajectories $\left \{ \phi^{(n)} _{21}, \phi^{(n)} _{31} \right \}$ of phase points ${\mathbf M}_{n} = \left (\phi^ {(n)} _{21}, \phi^{(n)} _{31} \right ) $ of the Poincar\'e map $\Pi:~ {\mathbf M}_n  \to {\mathbf M}_{n+1}$ are defined through the time delays $
\Delta \phi^{(n)} _{j1} = \displaystyle{\frac{\tau^{(n+1)}_{j1} - \tau^{(n)}_{j1}}{\tau^{(n+1)}_{1} -\tau^{(n)} _{1}}}
$ (on $\mbox{mod 1}$) between the burst initiations in each cycle normalized over the network period, can converge to several co-existing stable fixed points, thus indicating the given network is multistable, or a single stable invariant circle wrapping around the torus that corresponds to a unique rhythmic outcome with periodically varying phase lags. These are attractors, single or multiple, of the return map on a 2D torus, which are associated with multifunctional or dedicated neural circuits, respectively (Fig.~5). The 2D return map, $\Pi:~ {\mathbf M}_n  \to {\mathbf M}_{n+1}$, for the phase lags can be written as follows:
\begin{equation}
\phi^{(n+1)}_{21}=\phi^{(n)}_{21}+\mu_1 f_1 \left (\phi^{(n)}_{21}, \phi^{(n)}_{31} \right ), \qquad \phi^{(n+1)}_{31}=\phi^{(n)}_{31}+\mu_2 f_2 \left (\phi^{(n)}_{21}, \phi^{(n)}_{31} \right )
\end{equation}
with $\mu_i$ representing the coupling strength, and $f_i$ being some undetermined coupling functions such that $f_1=f_2=0$ corresponds to its fixed points: $\phi^{*}_{j1}=\phi^{(n+1)}_{j1}=\phi^{(n)}_{j1}$. These functions, similar to phase-resetting curves, can be assessed from the simulated data collected for known all trajectories $\left \{ \phi^{(n)} _{21}, \phi^{(n)} _{31} \right \}$. By treating $f_i$ as partials $\partial F / \partial \phi_{ij}$, we can restore a ``phase potential'' $F \left (\phi_{21}\, , \phi_{31} \right)=C$ that determines the dynamics of the coupled neurons, find its critical points associated with FPs -- attractors, repellers and saddles of the map, and by scaling $f_i$  predict their bifurcations due to loss of stability, and hence transformations of rhythmic outcomes of the network as a whole. 

\begin{figure}[h!]
\centering
{\includegraphics[width=0.98\textwidth]{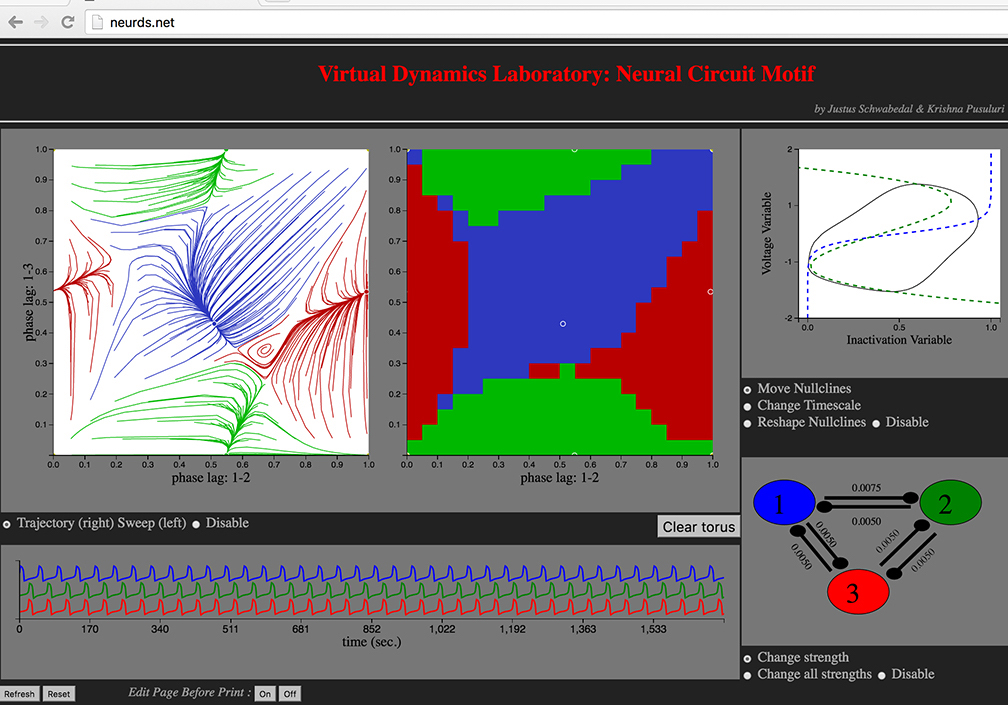}}
\caption{GPU-based interactive motif-toolbox \cite{justus,motif} for computational studies of rhythmogenesis in 3-cell circuits comprised of synaptically coupled FitzhHugh-Nagumo, Hodgkin-Huxley, and $2\Theta$-neurons, which can generate up to 6 (3 in this figure) robust patterns corresponding to the stable fixed points in the 2D Poincar\'e return map for the phase lags between constituent cells.} \label{fig5}
\end{figure}  

With such return maps, we can predict and identify the set of robust outcomes in a CPG with mixed, inhibitory and excitatory, slow or/and fast synapses, which are differentiated by phase-locked or periodically varying lags corresponding, respectively, to stable fixed points and invariant circles of the return map.  The toolkit lets us predict bifurcations and transformations of rhythmic outcomes before they actually occur in the network. The approach also reveals the capacity of the network and the dependence of its outcomes on coupling strength, wiring circuitry, and synapses, thereby letting one quantitatively and qualitatively identify necessary and sufficient conditions for rhythmic outcomes to occur. 
Using graphics processor units (GPUs) for parallel simulations of multistable neural networks using multiple initial conditions (as depicted in Fig.~5)   can drastically speed up the bifurcation analysis and reduce a simulation time to merely few seconds.   \\

\section{Acknowledgements} This work was funded in part by the NSF grant IOS-1455527 and the RSF grant 14-41-00044 at Lobachevsky University of Nizhny Novgorod. We thank the Brains and Behavior initiative of Georgia State University for providing pilot grant support. We acknowledge the support of NVIDIA Corporation with the Tesla K40 GPUs used in this study. Finally, we are grateful to all the current and past members of the Shilnikov NeurDS lab for productive discussions.

\bibliographystyle{unsrt}

\end{document}